\documentclass[10pt,conference]{IEEEtran}
\IEEEoverridecommandlockouts
\usepackage{cite}
\usepackage{amsmath,amssymb,amsfonts}
\usepackage{algorithmic}
\usepackage{graphicx}
\usepackage{textcomp}
\usepackage{xcolor}

\usepackage{array}
\usepackage{wrapfig}
\usepackage{multirow}
\usepackage{tabularx}
\usepackage{tcolorbox}

\usepackage[numbers]{natbib}

\begin{document}

\title{API2Com: On the Improvement of Automatically Generated Code Comments Using API Documentations
\thanks{This research is support by a grant from Natural Sciences and Engineering Research Council of Canada RGPIN-2019-05175.}
}

\author{\IEEEauthorblockN{Ramin Shahbazi}
\IEEEauthorblockA{\textit{Department of Computer Science} \\
\textit{University of British Columbia}\\
Canada, Kelowna \\
ramin20@mail.ubc.ca}
\and
\IEEEauthorblockN{Rishab Sharma}
\IEEEauthorblockA{\textit{Department of Computer Science} \\
\textit{University of British Columbia}\\
Canada, Kelowna \\
rishab.sharma@alumni.ubc.ca}
\and
\IEEEauthorblockN{Fatemeh H. Fard}
\IEEEauthorblockA{\textit{Department of Computer Science} \\
\textit{University of British Columbia}\\
Canada, Kelowna \\
fatemeh.fard@ubc.ca}

}

\maketitle

\begin{abstract}

Code comments can help in program comprehension and are considered as important artifacts to help developers in software maintenance.
However, the comments are mostly missing or are outdated, specially in complex software projects. 
As a result, several automatic comment generation models are developed as a solution. 
The recent models explore the integration of external knowledge resources such as Unified Modeling Language class diagrams to improve the generated comments.
In this paper, we propose API2Com, a model that leverages the Application Programming Interface Documentations (API Docs) as a knowledge resource for comment generation. The API Docs include the description of the methods in more details and therefore, can provide better context in the generated comments. 
The API Docs are used along with the code snippets and Abstract Syntax Trees in our model.

We apply the model on a large Java dataset of over 130,000 methods and evaluate it using both Transformer and RNN-base architectures. 
Interestingly, when API Docs are used, the performance increase is negligible. We therefore run different experiments to reason about the results. For methods that only contain one API, adding API Docs improves the results by 4\% BLEU score on average (BLEU score is an automatic evaluation metric used in machine translation). However, as the number of APIs that are used in a method increases, the performance of the model in generating comments decreases due to long documentations used in the input.
Our results confirm that the API Docs can be useful in generating better comments, but, new techniques are required to identify the most informative ones in a method rather than using all documentations simultaneously. 


\end{abstract}



\section{Introduction}


Code comments present a clear picture of source code using natural language summaries and can help developers understand software programs quickly \cite{ko2006exploratory, latoza2006maintaining}. 
Comments can play a major role in program comprehension to perform software maintenance and can reduce the time spent on understanding source code \cite{sridhara2010towards, xia2017measuring}. 
However, as software evolves, the comments are usually missed or are outdated \cite{moreno2013automatic, singer2010examination}.
As a result, there have been numerous research on building models that generate code comments automatically \cite{sridhara2010towards, eddy2013evaluating, haiduc2010use, zhu2019automatic}. 
In recent years, a main trend in automatic comment generation leverages neural networks \cite{iyer2016summarizing, hu2018deep, wan2018improving, movshovitz2013natural, bahdanau2014neural, haije2016automatic}. 
In these studies, the code comment generation is considered as a neural machine translation task in which the machine translates a code snippet written in a programming language such as Java to a piece of text written in English \cite{leclair2019neural}.

A main difference of the comment generation studies is the input used to feed the models : flattened Abstract Syntax Tree (AST) \cite{hu2018deep}, input code snippet \cite{iyer2016summarizing}, or leveraging external knowledge \cite{wei2019retrieve}. 
Recent studies show that when external knowledge is used, the quality of the generated comments increases \cite{wei2019retrieve, zhang2020retrieval, wang2020cocogum, haque2020improved}. 
For example, Zhang et al. retrieve similar code and comments to generate text for an unseen code snippet \cite{zhang2020retrieval}. 
Wang et al. incorporate class names as the intra-class context and Unified Modeling Language (UML) class diagrams as inter-class context. These two with code snippet and AST are used to generate comments \cite{wang2020cocogum}. 
The results of these and other works emphasize on using external knowledge sources to improve the quality of the generated comments.

\begin{figure*}
  \includegraphics[width=\textwidth]{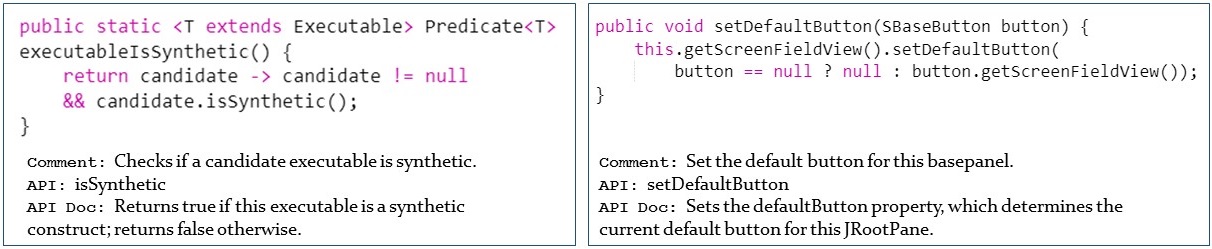}
  \caption{ Java methods, their comments, and the API documentation for the APIs used in the methods}
   \label{fig:motivation}
\end{figure*}

Application Programming Interface (API) documentations are fruitful resources for software developers and can be a candidate for being an external knowledge source for comment generation task. The API Documentations (API Docs) include descriptions about the methods and interfaces which can be beneficial for comment generation.  
As APIs are used frequently to implement code, we hypothesize that we can generate better comments to define the functionality of a code snippet when using the documentations of the APIs employed within the given method. 
For example, a common way to create and write to a JSON file is to use \verb|FileWriter.write| and 
\verb|FileWriter.flush|. 
More comprehensive comments can be generated by referring to their descriptions on the documentations \footnote{https://docs.oracle.com/javase/7/docs/api/java/io/OutputStreamWriter.html}: \textit{''writes a single character or a portion of a string"} and \textit{''flushes the stream"}, respectively. 
Another example is shown in Fig. \ref{fig:motivation} for two Java methods. 
The \textit{Comment} shown below each method is the text written by human taken from our dataset. In addition, the name of the API and its Documentation are shown below each one. 
The left method in this figure checks whether the candidate executable is synthetic, which is represented by the comment. When we consider the documentation of the \verb|isSynthetic| API, the \textit{returns false otherwise} can be added to the comment to complement it. 
Similarly, the API documentation of the method (extracted from JDK reference documentation\footnote{https://docs.oracle.com/en/java/javase/15/docs/api/index.html}) shown in the right of Fig. \ref{fig:motivation} can add more context to the comment. 
These documentations can form the functionality of the methods. 

Although the name of the API has been used in previous works \cite{hu2018summarizing}, the descriptions of the APIs have not been explored yet. 
Only using the API name may not be able to deliver the functionality of code properly. 
In contrast, the API documentation is based on natural language and includes more details, which can be useful for generating comprehensive comments. 
Therefore, we develop API2Com, a model that leverages the API Docs to enrich generating comments. In our model, we use API documentation, AST, and source code as input to a three-encoder architecture.
AST captures the structural representation of code and is previously used in several related studies \cite{hu2018deep, alon2018code2seq, leclair2019neural, wang2020cocogum, wei2019retrieve, zhang2020retrieval}. 
API Docs are added as another resource along with the code sequences to generate comments. 
We initially exploit a Transformer architecture \cite{vaswani2017attention} to learn the semantic representation of code. The reasons of choosing Transformer over other architectures are explained in Section \ref{sec:approach}. 
We conduct our study on a large Java dataset containing 137,007 records collected by Husain et al. \cite{husain2019codesearchnet} and run several experiments to understand the effect of adding API documentation to the model. 

Interestingly, although we use an external knowledge source, our results show that performance increase is negligible. Therefore, we conducted more experiments to reason about the results and report our findings in this paper. 
We find that as the number of APIs used in a method increases, the value of the API documentation decreases. 
This is mainly due to the fact that when more API Docs are used, a long text is fed to the model. However, the number of common words between the comments and the API Docs is extremely low. 
This causes the API documentation to add noise when the number of API Docs is more than three. 
We achieved similar results using Gated Recurrent Unit (GRU) architecture, the architecture that is used in many of the previous studies and is more efficient than the other gated mechanism approach, LSTM \cite{chung2014empirical}.
These findings show that API Docs can help in improving the comments, but new techniques should be developed to include only the informative ones. 
These results and our insights may save the researchers from navigating the same approach, while opening new avenues of research about integrating API documentations for comment generation.

The contributions of our study are summarized below:
\begin{itemize}
    \item We propose a Transformer model that combines API documentation with source code and AST to generate comments. For this purpose, we extract all the corresponding comments
    for the APIs used in a code snippet. 
    \item We perform several experiments in addition to comparing API2Com to other baselines to understand the effects of adding API Docs.
\end{itemize}

The rest of this paper is organized as follows. 
In Section \ref{sec:approach} we explain the details of our approach followed by the experiments and results in Sections \ref{sec:experiments} and \ref{sec:results}. 
Section \ref{sec:discussions} discusses the results. 
Threats to validity are mentioned in Section \ref{sec:threats}. A summary of the related studies are presented in Section \ref{sec:relatedWork} and we conclude the paper in Section \ref{sec:conclusion}.

\begin{figure*}
  \includegraphics[width=\textwidth]{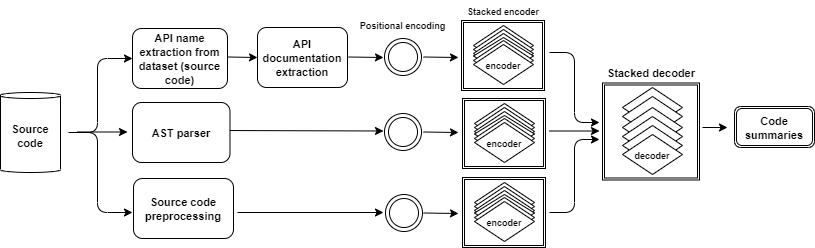}
  \caption{ An overview of API2Com architecture}
   \label{fig:Trans}
\end{figure*}

\section{Proposed approach} \label{sec:approach}

In our work, we exploit a Transformer architecture \cite{vaswani2017attention} to learn the semantic representation of code for the following reasons.
Most of the current studies use a Recurrent Neural Network (RNN) architecture such as Long Short Term Memory (LSTM) \cite{hochreiter1997long, wei2019retrieve, zhang2020retrieval} or Gated Recurrent Unit (GRU) \cite{leclair2019neural} for comment generation. 
RNN-based architectures have two limitations in capturing the representation of the source code. 
First, as they analyze the tokens sequentially, they fail to model the non-sequential code structure \cite{ahmad2020transformer}. 
Second, the RNN-based models are not able to capture the long-range relationship between words when the input is very long \cite{ahmad2020transformer}. 
On the other hand, Transformer can capture these relationships by adopting a self-attentional mechanism. 
Recently, Transformers are used in software engineering studies \cite{wang2019learning, you2019improving, fan2018hierarchical}, and proved to be effective for comment generation compared to other baselines \cite{ahmad2020transformer, wang2020trans}. However, these works use the code sequence as is, without incorporating external knowledge, thus incorporating API documentation in Transformer is unknown. 
Therefore, although we report the results of adding API Docs in an RNN architecture as well, we explain API2Com using a Transformer architecture in the following.

\subsection{Model overview}

The API2Com architecture is shown in Fig. \ref{fig:Trans}.
API2Com consists of three encoders, fed with code tokens, AST traversal sequence, and a list of API documentations, respectively. 
AST represents the syntactical structure of the code and is flattened by tree traversal to be fed to the model.
API Docs are in natural language and are complementary of the code semantics. Based on the capabilities of the architecture, source code represents both syntactical and semantic representation, which is enriched by AST and API Docs. 
The three encoders have similar architecture and their results are concatenated and passed to the decoder to generate comments.

\subsection{Transformer}

RNN-based neural network translators mostly exploit an encoder-decoder architecture where the encoder maps the input sequence to a vector, and the decoder uses the encoder's vector to generate the output sequence one word at a time. 
In our work, Transformers are used in encoders and decoder, which utilizes multi-headed self-attention mechanism and positional encoding. Details are explained below. 

\textbf{Encoder:} Encoder combines multiple identical layers where each layer consists of two sub-layers \cite{vaswani2017attention}. The first sub-layer forms a multi-headed self-attention structure while the other one is a fully connected layer. Both sub-layers are followed by another layer which normalizes the output of each sub-layer.

\textbf{Decoder:} Decoder has similar structure as encoder except that it includes one additional sub-layer per each stacked layer. This extra sub-layer conducts multi-head attention on the encoder's output \cite{vaswani2017attention}. Moreover, the self-attention sub-layer is reformed to avoid attending to subsequent positions.

\textbf{Multi-head attention mechanism:} Transformer's attention aims to map a query (Q) and a collection of key(K)-value(V) pairs to vectors which are calculated as a weighted sum of the values. The attention is computed by multiplying queries by keys, divided by $ \sqrt{d_{k}} $, where $d_{k}$ is the dimension of the key vector. Then, a softmax function is applied to attain the weights for the values. 

\begin{equation}
    Attention(Q,K,V)=Softmax(QK^T/\sqrt{d_{k}})V 
\end{equation}

In order to improve the performance, a new mechanism called multi-head attention is added to the self-attention technique. The intuition is to conduct self-attention process multiple times separately, with different weight matrices. However, the feed-forwarding layer only accepts a single input. Therefore, all the results are concatenated and multiplied by an additional weight matrix $W$, as shown in the following formula \cite{vaswani2017attention}. 
Here, $d$ represents the dimension of the vectors.

\begin{equation}
\begin{aligned}
     MultiHead(Q,K,V) = Concat(head_{1},\\ head_{2}
    , ..., head_{8} ) W^o
\end{aligned}
\end{equation}

\begin{equation}
\begin{aligned}
     head_{i} = Attention(QW_{i}^Q,KW_{i}^K,V_{i}^V) 
\end{aligned}
\end{equation}

Where $W^Q_{i}\in R^{d_{model} * d_k   }$ ,
$W^K_{i}\in R^{d_{model} * d_k   }$,
$W^V_{i}\in R^{d_{model} * d_v   }$ 

RNN-based models are incapable to capture the relationship between tokens effectively when the input sequence is long while integrating a multi-head attention mechanism with transformer addresses this issue.

\textbf{Positional Encoding:} Given that the model does not include recurrence, some information regarding relative and absolute tokens' positions needs to be taken into account. To achieve this goal, a positional encoding layer is added at the lowest part of the encoder and decoder stacks. 
The layers should have equal dimensions as the output results of embedding should be added to positional encodings. In this technique \cite{shaw-etal-2018-self}, the relationship between two tokens $i$ and $j$ are shown by vectors $a_{ij}^V, a^K_{ij} \in  R^{d_k}$. The representation of relative position is included in the output of each sub-layer which is the input to the next layer. Every head calculates the following weighted sum:

\begin{equation}
\begin{aligned}
    z_i = \sum_{j=1}^{n} a_{ij} (x_j W^V + a_{ij}^V)
\end{aligned}
\end{equation}

Where $x_j$ is the state of the encoder, $z_i$ is the output of the attention and $a_{ij}$ is calculated by a softmax function.
Shaw et al. \cite{shaw-etal-2018-self} supposed that relative position information is not effective enough in long-distance cases. Therefore, a constant parameter $k$ is defined as maximum relative position. With the assumption that preceding tokens are in a negative direction and succeeding tokens are in a positive direction, the relationship of relative position between two words is computed by $2k+1$ unique labels as below, where $w^K$ and $w^V$ are the position representations.

\begin{equation}
\begin{aligned}
    a_{ij}^K = w^K_{clip(j-i,k)}; 
     a_{ij}^V = w^V_{clip(j-i,k)}\\
     clip(x,k) = max(-k, min(k,x))
\end{aligned}
\end{equation}

\subsection{API documentation}

We extract the API documentation knowledge for every API used within each method from JDK reference documentation. JDK reference documentation contains documentation for various levels including packages, classes, and methods, and we only include the methods' documentation in API2Com. 
Consider the method used in the left part of Fig. \ref{fig:motivation}. The method includes one API named \verb|isSynthetic()| which is listed as a method under the class \verb|java.lang.reflect.Executable| in JDK documentation. 
The documentation of this class contains tables for summary of methods which has two fields: i) Modifier and Type and ii) Method and Description. We extract the \textit{description} of the method. In this example, the text \textit{''Returns true if this executable is a synthetic construct; returns false otherwise."}\footnote{https://docs.oracle.com/javase/8/docs/api/java/lang/reflect/Executable.html} is extracted. 
These documentations for all of the APIs within a method are extracted and after concatenation are used as a single input to the first encoder of the model.

\section{Experiments} \label{sec:experiments}

In this section, the details of the experimental setup are presented.

\subsection{Dataset and Preprocessing}

We use the Java portion of the CodeSearchNet dataset, which is introduced by Husain et al. \cite{husain2019codesearchnet}, to train and test API2Com. 
This dataset is used in recent studies for various software engineering tasks including comment generation \cite{feng2020codebert,wang2020cocogum} and has high quality.
CodeSearchNet is a set of datasets that were initially collected by scraping open-source GitHub repositories with the assistance of Libraries.io. 
This tool aids identify the projects which have already been used by other projects with the purpose of extracting ones with higher validity. 
The methods and their associated comments are included in the dataset. 
The authors have removed all the projects without a valid license. 
They also excluded all comments that have less than three tokens and all implementations that are less than three lines to ensure about the quality of the dataset. 
The constructors, extension methods, and methods with the word ''test" within their names, duplicates and auto-generated methods are also eliminated from the dataset. 
These steps are conducted by the publishers of the dataset \cite{husain2019codesearchnet} to increase the quality of the dataset by removing biases for training neural models for source code representation.

We set the code and comment length to 256 and 64 respectively, similar to the work of Feng et al. \cite{feng2020codebert}.  
As AST cannot be produced for truncated code and it might cause loss of valuable information, we have to exclude the records with the code or comment length higher than 256 or 64. This step is required to have a fair comparison between all models, as some models use AST. 
Table \ref{table:Datasetstatistics} shows the statistics of the dataset.  The dataset is divided into train, validation and test sets with 8/1/1 proportions, respectively. 
We use the train and validation sets to train our model, and the test set is used in our evaluations. 
Following previous works \cite{leclair2019neural}, we preprocess the dataset further before training by splitting the tokens into sub-tokens where CamelCase or snake\_case is used; converting all tokens to lower case; and removing punctuations from code.

\begin{table}[h!]
\caption{Dataset statistics}
\label{table:Datasetstatistics}
\small	
\begin{tabular}{ p{0.2\linewidth}p{0.3\linewidth}p{0.3\linewidth} }
 \hline
Dataset & Number of functions before filtering  & Number of functions after filtering\\
\hline
Train & 164,923 & 137,007 \\
Validation &	5,183 & 4,326    \\
Test &	10,955 &	9,011\\

 \hline
\end{tabular}
\end{table}

\subsection{API Documentation Extraction}

We developed a scraper to download all the APIs and their corresponding descriptions within every module and class in JDK reference documentation. 
To extract the API names from the methods in our dataset, we utilized srcML tool\footnote{https://www.srcml.org/}.
srcMl is a tool that can build AST tree from source code and present the results as XML tags. This representation provides information about the tokens. For example, it can label a token as 'methodCall' if it is calling another function. Using this information, we are able to extract all APIs names by traversing the AST tree. 
We then match the API names used in each method to the scraped dataset to retrieve their descriptions. 
For overloaded functions, we use the number of parameters to extract the correct documentation. 
If there are multiple occurrences of the same API with different API Docs in the JDK reference documentation, we use the documentation which has highest frequency. 
For example, \texttt{toString()} is a method that is used in 455 Java classes/interfaces. For these cases, the API Docs for all of them are extracted. 
We then group the ones for which the text is exactly the same. As the information from parent class is not included in the AST, we use the API Doc from the largest group as it is the most frequently used text.

\subsection{Model Training}
To train the model, we use the API Docs in the first encoder, flattened AST is the second one, and the pre-processed source code in the last encoder. The results of these three encoders are concatenated to pass on to the decoder as a single input.

\subsection{Experiment setting}

We develop the model with PyTorch framework\footnote{https://pytorch.org/}. The number of encoder and decoder layers is set to 6, and the number of heads in the multi-head sub-layer is set to 8. \textit{stochastic gradient descent} optimizer with an initial rate of $0.1$ is used. We replaced the out-of-vocabulary tokens with UNK. 
The dimension of the hidden state and batch size is set to 512 and 32 respectively. 
To reduce the probability of overfitting, dropout with the value of $0.1$ is used. 
Training is executed for $100$ epochs, although it would stop if learning rate decay to $10^{-7}$ . The best epoch is selected based on the value of the loss function for the validation dataset. We perform the experiments on a Linux server with the NVIDIA Tesla V100 GPU with 32 GB memory.

\subsection{Evaluation Metrics}
Similar to previous studies in code comment generation \cite{ahmad2020transformer, wan2018improving, zhang2020retrieval}, we evaluate the performance based on the following metrics,
BLEU \cite{papineni2002bleu}, ROUGE-L \cite{lin2004rouge}, and METEOR \cite{banerjee2005meteor}. These scores are evaluations metrics used to assess the ability of the model in predicting text similar to the comments written by developers for the given code snippets (i.e. Reference text).
BLEU-n score quantifies the average n-gram accuracy between the reference and predicted sentences, with the assistance of a brevity penalty \cite{papineni2002bleu}, and is reported for $n \in [1,4]$. 
ROUGE-L utilizes F-score which is computed as a weighted harmonic mean of recall and precision values obtained from finding the longest common sequence of the texts \cite{lin2004rouge}. 
METEOR, is a recall-based metric which gauges how the model performs in capturing the content of reference sentences and is based on the number of identical n-grams between reference and predicted texts \cite{banerjee2005meteor}.

\subsection{Baselines} \label{baselines}

\begin{table*}[h!]
\caption{BLEU, METEOR, and ROUGE-L scores of API2Com and other baselines}
\label{table:results}
\small	
\begin{tabularx}{\textwidth}{ p{0.21\textwidth}p{0.1\textwidth}p{0.1\textwidth}p{0.1\textwidth}p{0.1\textwidth}p{0.1\textwidth}p{0.12\textwidth}  }
 \hline
 Models &\multicolumn{6}{c}{Metrics} \\
 \cline{2-7}
 & BLEU-1(\%) & BLEU-2(\%) & BLEU-3(\%) & BLEU-4(\%) & METEOR(\%) &  ROUGE-L(\%) \\
 \hline
 AST-Attendgru   &22.02  &9.97  &5.21  &3.20  & 9.45 & 23.06\\
 TL-CodeSum&   21.65  & 10.10   &5.69&3.74&8.36&22.10\\
 Rencos &22.82  & 12.29&  7.54 &5.15&10.55&\textbf{27.46}\\
 \hline
 Transformer-based    &\textbf{25.53} & \textbf{16.27}&  \textbf{10.68} &\textbf{7.33}&\textbf{14.07}&\textbf{31.97}\\
 \hline
 \hline
 API2Com\textsubscript{Base}&   24.62  & 13.80&8.22&5.27&10.78&25.78\\
 API2Com\textsubscript{AST}& 24.52  & 13.59&8.13&5.26&10.74&25.62\\
 API2Com\textsubscript{API} & \textbf{24.78}  & \textbf{13.87}  & \textbf{8.26} &\textbf{5.28} &\textbf{10.85}&25.88\\
  API2Com\textsubscript{Full} & 24.69 &	13.85&	8.24&	5.26&	10.79&	25.80  \\
 
 \hline
\end{tabularx}

 \end{table*}

We evaluate the effectiveness of API2Com based on the existing state-of-the-art studies for comment generation. We ran each of the baselines on the CodeSearchNet dataset to compare the result with our approach. Effort has been taken to use the best hyperparameters identified by the authors of each model.

\textbf{AST-Attendgru} \cite{leclair2019neural} leverages an attentional GRU encoder-decoder model to generate code summaries in natural language. It integrates AST traversal sequence, flattened by Structure Based Traversal technique, into the model with the assistance of a second encoder to capture the structural information of the source code more effectively.

\textbf{TL-CodeSum} \cite{hu2018summarizing} initially trains an RNN encoder-decoder model on API sequence and comment pairs, trying to build a mapping between API knowledge and method comments. They further transfer the trained encoder, as a second encoder, to another Seq2Seq model which receives code token sequence along with API sequence extracted from source code to generate summaries. In other words, the learned encoder which is a representation of API sequence is employed in the task of code comment generation.

\textbf{Rencos} \cite{zhang2020retrieval} is an attentional LSTM seq2seq model. The authors enhance the model by retrieving the two most similar methods to the input code from the training set in terms of both semantic and syntactic similarity. The model receives the input code and its similar methods and generates comments by fusing the inputs in the decoding process.

\textbf{TransformerBased} is one of the first studies that utilizes Transformer for comment generation. Ahmad et al. \cite{ahmad2020transformer} investigate the effectiveness of the transformer model for generating code comments. They further incorporate relative positional encoding and copy attention mechanism into the transformer to improve the quality of generated comments.

\textbf{Variations of API2Com} We use variations of API2Com to compare the effectiveness of its components. API2Com\textsubscript{Base} is the Transformer architecture without using AST or API Docs. API2Com\textsubscript{AST} uses two encoders to encode AST and input code, without using API Docs. Similarly, API2Com\textsubscript{API} only uses input code and API Docs. Finally, API2Com\textsubscript{Full} is the model which uses all three inputs: code, AST, and API Docs. 
We also employ API2Com using an RNN-based architecture with same variations and refer to it as API2ComS and discuss the results in the second research question.

\section{Results} \label{sec:results}

\subsection{Research Questions}

In this section, we present results of our experiments.

\begin{table*}[h!]
\caption{Evaluation metrics for the four variations of RNN-based API2Com (API2ComS)}
\label{table:seq2seqresult}
\small	
\begin{tabularx}{\textwidth}{ p{0.19\textwidth}p{0.1\textwidth}p{0.1\textwidth}p{0.1\textwidth}p{0.1\textwidth}p{0.1\textwidth} p{0.12\textwidth} }
  \hline

 Model & BLEU-1(\%) & BLEU-2(\%) & BLEU-3(\%) & BLEU-4(\%) &   METEOR(\%)  & ROUGE-L(\%)\\
 
 \hline
 API2ComS\textsubscript{Base}&   20.75&	9.36	&5.23&	3.68 & 8.24 &	21.99\\
 API2ComS\textsubscript{AST}& 20.59 &	9.31 &	5.21 &	3.67 &  8.26  &	21.81\\
 API2ComS\textsubscript{API} & 20.86 &	9.41 &	5.27 &	3.69 & 8.29   &	22.13\\
 API2ComS\textsubscript{Full} & 20.73  & 9.34	&5.23	& 3.67	& 8.27 &	21.95  \\
 
 \hline
\end{tabularx}

 \end{table*}

\textbf{RQ1: How does our proposed approach perform compared to the baselines?}
Table \ref{table:results} shows the evaluation metrics of API2Com and other baselines on the test set.
Among all models, the Transformer-based achieves the best performance for all metrics. 
This model is the only model from baselines that leverages the Transformer architecture. 
Transformer-base uses advanced techniques and varies from the original Transformer architecture that we use in API2Com. 
We relate the best results of this model to incorporating relative positional encoding and copy mechanism. 

The other baselines use RNN-based architectures, which all have lower scores compared to the Transformer models. Transformer is the state of the art architecture used in neural machine translation which processes the input non-sequentially using self attention and positional encoding. These mechanisms allow Transformers to learn the relational information between words. 
Interestingly, adding AST to Transformer decreases the scores slightly. This is seen in our model API2Com\textsubscript{AST} compared to API2Com\textsubscript{Base}, which acknowledges the strength of this architecture in capturing the relationship between input sequence tokens.

The second best result belongs to API2Com\textsubscript{API} for all metrics except ROUGE-L. After Transformer-base and API2Com, Rencos has the best results, followed by AST-Attendgru and TL-CodeSum. 
TL-CodeSum uses the API names as a knowledge source to enhance the generated comments, but still has lower results compared to API2Com and Rencos. 
In some scores, AST-Attendgru achieves higher performance and in some other, TL-CodeSum.
Note that CodeSearchNet data is a large curated dataset. A specific feature of this dataset is having longer methods in which other methods are used within. This might have an effect on having lower results for the baselines, compared to results published by their authors. However, when Smooth-BLEU score \cite{wang2020cocogum, feng2020codebert} is applied, we achieve similar results to the ones reported in \cite{wang2020cocogum} which uses the same dataset. This ensures the correctness of our results.

\textbf{RQ2: What is the effect of each component of API2Com in generating comments? }
We evaluate the effect of adding AST and API documentations to the Transformer model. For this purpose, we train different variations of our model as shown in Table \ref{table:results}. 
In addition, we evaluate our model on a RNN-base architecture, which uses GRU architecture instead of a Transformer architecture. 
We refer to this model as API2ComS and show four variations as API2ComS\textsubscript{Base}, API2ComS\textsubscript{AST}, API2ComS\textsubscript{API}, and API2ComS\textsubscript{Full} to refer to the model when only code is used in the input, code and AST are used, code and API Docs are used, or all three inputs are used to train the model. 
This comparison can provide more insights about the effect of each component independent of the architecture used. 
The results of Seq2Seq variations are shown in Table \ref{table:seq2seqresult}, demonstrating that in either architecture, AST decreases the performance and incorporating API documentations increases the scores. 
However, the change in the score in both cases is very small. 
Using the Full model also only has a slight change in all scores. 
Note that the results of API2ComS is a bit lower than TL-CodeSum. We relate this to the way TL-CodeSum is trained as explained in section \ref{baselines}. Although we could use similar knowledge transfer for API2ComS, we kept the model similar to API2Com for fair comparison.
Similar to the results of RQ1, the Transformer-based API2Com has higher scores than its RNN-based model, API2ComS. 
As AST reduces performance in our model, for the rest of the paper we conduct experiments on API2Com\textsubscript{Base} and API2Com\textsubscript{API}.

\begin{table*}[h!]
\caption{Results of API2Com\textsubscript{Base} and API2Com\textsubscript{API} on separated datasets based on frequency of APIs used in methods}
\label{table:API-Table}
\small	

\begin{tabularx}{\textwidth}{p{0.13\textwidth} p{0.16\textwidth}p{0.07\textwidth}p{0.07\textwidth}p{0.07\textwidth}p{0.07\textwidth}p{0.07\textwidth}p{0.07\textwidth}p{0.08\textwidth}  }
 \hline
 API Frequency & Model& \multicolumn{7}{c}{Metrics} \\
 \cline{3-9}
 && BLEU-1 & BLEU-2 & BLEU-3 & BLEU-4 & METEOR & ROUGEL & Avg. \% \\
 
 \hline
1 API & 
API2Com\textsubscript{Base}&   20&9.71&4.67&2.43&8.1&20&
\\
 &API2Com\textsubscript{API} & 20.67&10.16&4.95&2.48&8.27&20.84&
\\
&Improvement (\%) & 3.35\%&4.63\%&6\%&2.06\%&2.1\%&4.2\%&3.72\%
\\
  \hline
2 APIs&
 API2Com\textsubscript{Base} & 19.24&9.36&4.35&2.05&7.6&20.58&\\
 &API2Com\textsubscript{API} & 19.55&9.52&4.53&2.11&7.82&20.83&\\
 &Improvement (\%) & 1.61\%&1.71\%&4.14\%&2.93\%&2.89\%&1.21\%&2.42\%\\

 \hline
 
 3 APIs&
 API2Com\textsubscript{Base} & 18.22&8.86&4.16&2.12&6.76&18.89&\\
 &API2Com\textsubscript{API} & 18.25&8.95&4.27&2.19&6.75&18.81& \\
 &Improvement (\%) & 0.16\%&1.02\%&2.64\%&3.3\%&-0.15\%&-0.42\%&1.09\%\\

 \hline
 
  More APIs&
 API2Com\textsubscript{Base} & 19.79&9.71&4.86&2.54&7.91&20.16& \\
 &API2Com\textsubscript{API} & 19.52&9.66&4.79&2.49&7.9&20.12& \\
 &Improvement (\%) & -1.36\%&-0.51\%&-1.44\%&-1.97\%&-0.13\%&-0.2\%&-0.94\% \\

 \hline
\end{tabularx}
\end{table*}

\textbf{RQ3: What is the effect of number of APIs on the performance of API2Com?}
In this research question, we investigate the performance of API2Com\textsubscript{API} in different subsets of our dataset: when the methods have one API, two APIs, three APIs, and four or more APIs.
Every method might include multiple APIs to perform its functionality. 
In API2Com, we concatenate the API Docs of the APIs used in a method and use it as a single input to the model. We suspect that this concatenation might be a potential reason for not having a significant improvement in API2Com\textsubscript{API}; 
as this process results in having multiple lines of documentation next to each other and might add noise to the comment. 
To verify this hypothesis, we conduct an experiment by splitting the dataset with regards to the frequency of APIs used in the methods. 
Table \ref{table:SimilarityTable} shows the average length of comments and the API Docs in each split. 
We train and test the API2Com\textsubscript{Base} and API2Com\textsubscript{API} models on the newly separated datasets. 
The results are presented in Table \ref{table:API-Table}. 
The last column of the table shows the average of the improvement scores for all metrics. 
The results confirm that adding API Docs is effective, although being small. However, as the number of APIs increases, this effect becomes less. 
Adding one API can improve the performance by $3.72\%$ on average, which decreases to $2.42\%$ for two APIs and $1.09\%$ for three APIs. 
This effect can be related to the length of API Docs added to the input which are shown in Table \ref{table:SimilarityTable}. 
When one API is used, the documentation has approximately the same length as the code comment, which approximately increases to two and three times of the comment length for 2 APIs and 3 APIs, respectively. 
When more than 3 APIs are used, the documentation length added to the input is more than 5 times of length of the code comment.
This increased length has a negative impact on the performance of the model in this case ($-0.94\%$).
The methods in the test set with more than three APIs are over 25\% of the whole test dataset. This large proportion can negatively affect the results reported in RQ1 and RQ2, and lead in negligible improvement for the results of API2Com\textsubscript{API} compared to API2Com\textsubscript{Base}.

\begin{table}[h!]
\caption{Comment and API documentation length}
\label{table:SimilarityTable}
\small	

\begin{tabular}{ p{0.27\linewidth}p{0.29\linewidth}p{0.29\linewidth} }
 \hline
Dataset & Comment Length & API Docs length\\
\hline

         Entire dataset&12.8 & 27.2 \\
         1 API&12.8 & 10.8\\
        2 APIs&12.9 & 21.8\\
        3 APIs&13.0 &32.3 \\
     More APIs&13.0 & 69.4\\
 \hline
\end{tabular}

 \end{table}

\subsection{Human Evaluation}

We evaluate the results by conducting a qualitative analysis. 
In this experiment, we sampled 100 records randomly, such that there are equal number of records for each API frequency used in RQ3. For example, there are 25 records with methods that contain three APIs. 
Following previous studies \cite{zhang2020retrieval}, we compare the ''Reference" text, and the comments of API2Com\textsubscript{Base} and API2Com\textsubscript{API}. 
For this purpose, we use 300 HITS using Amazon Turk\footnote{https://www.mturk.com/}. Each sample is evaluated by 3 random subjects and finally the average score over each group with the same number of APIs is calculated. In total, 43 evaluators have participated in this survey. The evaluators are asked to give a similarity score between 1 to 5 to each of the generated comments compared to the Reference text. Here, 1 shows the lowest similarity and 5 represents the highest similarity.  

The average score obtained by API2Com\textsubscript{Base} and API2Com\textsubscript{API} over all samples (not separated based on the number of APIs) are $2.99$ and $3.16$, respectively. This result confirms that similar to the automatic scores, the API2Com\textsubscript{API} improves the generated comments, but the improvement is small. 
The average scores when the samples are separated based on the number of APIs are similar to the results explained in RQ3. When one API is used, the difference between the two models are larger and API2Com\textsubscript{API} rises the score from $3.08$ to $3.39$, and adding more APIs reduces the improvement. 
In contradiction to the automatic metrics, when more than 3 APIs are used, the average score by human evaluation for API2Com\textsubscript{API} is $3.17$, higher than the $2.73$ score for API2Com\textsubscript{Base}. The reason might rely in the fact that for human evaluation, the comments are generated by the model which is not separated based on API frequencies, but in RQ3, the models are trained separately for each dataset containing different number of APIs. We did not separate the models here, as we are interested in evaluating the API2Com\textsubscript{API} in general.

\section{Discussions} \label{sec:discussions}

\begin{figure*}
  \includegraphics[width=\textwidth]{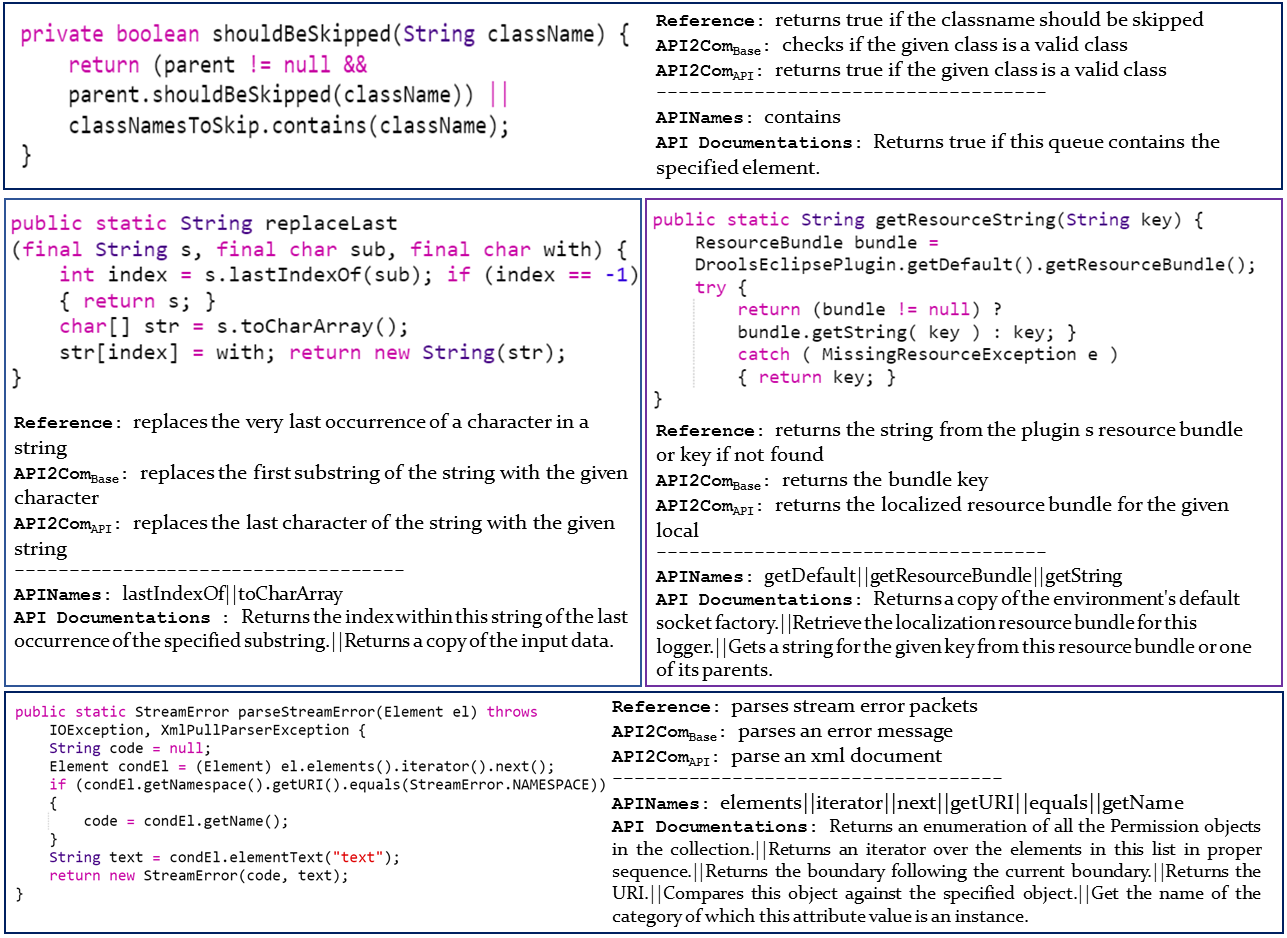}
  \caption{Java methods with different number of APIs. As the number of APIs increases, the effect of adding API Docs in creating better comments decreases.}
   \label{fig:api-numbers}
\end{figure*}

\textbf{Number of API Docs: }
We discussed that adding API documentations is improving the generated comments, as long as the number of documentations concatenated is less than three. 
Fig. \ref{fig:api-numbers} shows four methods, and their "Reference" comment from the dataset, the comments generated by API2Com\textsubscript{Base} and API2Com\textsubscript{API}, name of APIs used in each method, and the concatenated API documentations. The top method has only one API and the API Doc helps in generating \textit{Returns true} in the comment. 
The left center method has two APIs. Although the documentation of the first API (\texttt{lastIndexOf}) helps in generating better comments, the other documentation does not add any information. 
The right center method contains three APIs and it seems that the API Docs in this case are adding noise to the model, therefore not helping in generating better comments. 
Finally, The bottom method has six APIs which is prompting the model to generate wrong comment. API2Com\textsubscript{Base} is generating \textit{"error message"} which is correct while the API2Com\textsubscript{API} is misled by the API documentations resulting in a wrong phrase \textit{"xml document"}.

\textbf{API categories: }
We conduct another experiment to investigate whether the category of the APIs can impact the obtained results, following the approach used in \cite{hu2020deep}. For this purpose, we randomly selected 368 methods, a representative sample size of the test set with confidence level of 95\% and confidence interval of five\footnote{https://www.surveysystem.com/sscalc.htm}. 
First, we selected 100 random samples (not common with the representative samples). Two of the authors with three years of experience in Java development independently read the 100 methods and identified the categories; where the categories are from Hu et al. and the authors assessed existence of new categories as well \cite{hu2020deep}. For example, they identify whether the given code is about File Operation or Object Creation. The authors then discuss their categories together to come up to a consensus. 
Then, one of them used this predefined category and assigned the 368 methods to each group. A third person who is a software developer with 5 years of experience double checked this grouping. 
The final set of categories and the number of methods in each one is presented in Table \ref{table:categories}. 
For each category, we calculate the BLEU scores for API2Com\textsubscript{Base} and API2Com\textsubscript{API} reported in the table. 
The scores for the Setting Update and Algorithm Implementation increase, while the scores of the Object Creation/Retrieve, String Process, and File Operation decrease when API Docs are added. This change can be related to the documentations in each category, being helpful in some categories and less informative in others. 

We also explored the samples manually to find a specific pattern, with the purpose of identifying whether the APIs in a specific category can help improve the results. However, we find contradicted examples in all categories, when adding more APIs helps in generating better comments or vice versa. 
We relate this discrepancy to two reasons. 
First, the automatic scores are not 100\% reliable, as they are based on the occurrence of words, and do not consider the semantic similarities of the texts \cite{stapleton2020human}. Second, the importance of the API Docs in generating better comments depends on the intended functionality of the method. An API documentation might be helpful in one case and does not add much information for another method. 
It is worth mentioning that more analysis is required to investigate the effect of the categories and the number of APIs together. Finding enough samples in each category and with different number of APIs is ambitious, as identifying the categories is non-trivial and requires accurate techniques if done automatically.

\begin{table}[h!]
\caption{Categories for 368 random sampled data and BLEU scores of API2Com\textsubscript{Base} and API2Com\textsubscript{API}}
\label{table:categories}
\small	
\begin{tabular}{ p{0.39\linewidth}p{0.06\linewidth}p{0.17\linewidth}p{0.17\linewidth} }
 \hline
Categories & Count & API2Com\textsubscript{Base} & API2Com\textsubscript{API}\\
\hline
Entire samples & 368 & 23.45 & 23.26\\
Setting Update &	79 &	27.05 &	27.81 \\
Object Creation/Retrieve &	87 & 23.57 & 21.59	 \\
Algorithm Implementation &	131 &	19.24 &	20.37 \\
String Process &	30 &	27.32 &	25.12 \\
File Operation &	41 &	26.99 &	26.03 \\

 \hline
\end{tabular}
\end{table}

\textbf{Low frequency words: }
One of our observations is that adding APIs can increase the number of low frequency words used in the generated comments. 
Table \ref{table:low-frequency-words} demonstrates the number of words that occur less than 500 times in the corpus and their frequencies in the reference comments in our dataset. The table also shows the number of words from each frequency group that appear in the generated comments by API2Com\textsubscript{Base} and API2Com\textsubscript{API}. 
To determine these numbers, we extract all the tokens that are generated by our model which exist in the reference comments. 
We then count the number of occurrences of every extracted token in the train set, and report their frequencies. We also count the number of tokens generated by our model in each of the frequency categories. 
Although in general the number of low frequency words captured by the pure Transformer architecture is low, adding API Docs can increase the percentage of appearance of these tokens in the comments by $2\%$ on average, compared to API2Com\textsubscript{Base}. 
Note that API2Com is not designed with the purpose of solving low frequency words' problem. However, it helps boosting the base model by $5\%$ for words with frequency less than 50 and generates $3\%$ of these words in the comments. For the words with frequencies above 50, their appearance in the comments of API2Com\textsubscript{API} is above $11\%$. 
This number increases gradually and is $19\%$ as the word occurrence increments to 400-500 times.

\begin{table*}[h!]
\caption{The number of occurrences of low frequency words in the comments generated by API2Com\textsubscript{Base} and API2Com\textsubscript{API}}
\label{table:low-frequency-words}
\small	
\begin{tabularx}{\textwidth}{ p{0.17\textwidth}p{0.05\textwidth}p{0.09\textwidth}p{0.1\textwidth}p{0.1\textwidth}p{0.1\textwidth}p{0.1\textwidth}p{0.1\textwidth}}
 \hline
& $<50$ & 50-100&100-150 & 150-200& 200-300&300-400& 400-500\\
\hline
Reference&6042&3060&2494&2289&3664&3059&2133\\
API2Com\textsubscript{Base}&175&352&326&376&584&575&395\\
API2Com\textsubscript{API}&184&366&332&384&576&573&406\\

 \hline
\end{tabularx}

 \end{table*}

\textbf{Adverse effect: }
During our observations, we have seen examples that adding more APIs helps the model generate better results. 
For example, \textit{''remove all occurrences of the character c at the end of s'} is a Reference comment for one of the methods in our dateset. This method removes a character from end of a string and uses four APIs \texttt{indexOf}, \texttt{length}, \texttt{charAt}, and \texttt{substring}. 
API2Com\textsubscript{Base} generates \textit{removes the end of the string} and API2Com\textsubscript{API} produces \textit{remove all occurrences of the given string}. 
Obviously, adding more APIs has helped the model to capture the word \textit{occurrences} which conveys the functionality of the method better. In this example, this is mainly due to the API Doc of \texttt{indexOf} described as \textit{''Returns the index within this string of the first occurrence of the specified substring."}.


\begin{tcolorbox}
As the higher number of APIs decreases the model's performance, designing a strategy to only retrieve the informative APIs might be a more effective approach. For example, \texttt{toString} with API documentation \textit{'return a string representation of the object'} is a common API used frequently in Java methods. However, it rarely might provide valuable information to explain the functionality of the given method. 
On the other hand, consider the \texttt{getResourceBundle} API in bottom right method of Fig. \ref{fig:api-numbers}. The documentation of this API can add more context to the generated comments. 
Thus, finding a technique to only extract the informative APIs might help capture the knowledge from API documentation and prevent adding noise.
\end{tcolorbox}

\section{Threat to validity} \label{sec:threats}

\textbf{Internal validity: } 
Our model does not check the differences between the APIs and considers the API Docs for all APIs used in a method.
This may result in having repetitive texts concatenated to each other when feeding the API encoder, which might have resulted in lowering the scores when more than three APIs are used. 
However, we have noticed examples with such repetitions for which API2Com\textsubscript{API} improved the scores by $33\%$ and generated comments similar to the Reference text. 
An example is when the Reference comment is \textit{''converts a rectangle into a polygon'} and API2Com\textsubscript{API} has generated \textit{converts the given polygon into a polygon}. 
This is notable that the word \textit{polygon} is a key concept in this Reference comment, which is not captured by other models. For example, Rencos creates \textit{converts a single one location in the directly none} as the comment for this example. 

Another consideration is that we do not use techniques such as beam search used in previous works \cite{ahmad2020transformer}, which helps in investigating different paths and produce better comments. This is due to being computationally costly, especially with Transformers. 
This can lead to some repetitive words occurring in the output of API2Com\textsubscript{API} in some cases. This might be a pitfall for low scores, but during our manual analysis, we did not find many occurrences of such repetitions.  
Another threat can be developing neural networks for our model and reproducing the results of baselines. We have double checked our code for correctness. For running the baselines, we used the public repositories of the models and used the best parameters suggested by authors to alleviate the threats. 

\textbf{External validity: } 
We use a large high quality dataset to increase the validity of the results for Java. 
However, the results might not be generalizable to other programming languages, as Java may not be a comprehensive representation of other programming languages. 
Though, API2Com can be applied to other languages as it is not language-specific. 

\textbf{Construct Validity: }
In all of our evaluations, we calculated the metrics used in comment generation works. To alleviate the potential issues proposed by one metric, we report the results for BLEU, METEOR, and ROUGE-L. In addition, we evaluated the results by human judgement, which ensures the comparison between texts generated by the base model and the model enhanced by API Docs.

\section{Related works}  \label{sec:relatedWork}

Code comment generation has a long history using templates based approaches \cite{sridhara2010towards, mcburney2015automatic} and techniques based on information retrieval \cite{haiduc2010use, wong2013autocomment}. 
Since 2016, many researchers use neural network frameworks for comment generation. Neural network models can outperform the traditional approaches with the existence of large scale datasets.  Iyer  et  al. \cite{iyer2016summarizing} first applied the encoder-decoder seq-seq model where the encoder only contains token embedding and the decoder consists of an LSTM layer. 
Hu  et  al. \cite{hu2020deep} proposed DeepCom which passes the source code to the encoder in a different way. They converted the source code to AST to capture the structural information of the source code. They further introduced a structure-based tree traversal for flattening the generated AST to sequential text and use it as input to the LSTM encoder.
LeClair et al. \cite{leclair2019neural} improved the performance of the code generation task by capturing both the semantic and structural information of the source code. They used source code as well as flatten AST as input to two GRU encoders, and then results are used in the decoder to generate text. 
Alon  et  al. \cite{alon2018code2seq} introduced a novel flattening technique to traverse AST and convert it to text sequence. They considered various traversal paths and used them as distinct inputs to the LSTM encoder-decoder model. 
Wei et al. \cite{wei2019code} consider the relation between code generation and comment generation to improve both tasks. They design a dual training model to incorporate duality on attention weights. 
Feng et al. \cite{feng2020codebert} propose using a natural language - programming language bimodal pre-trained model \cite{devlin2018bert, liu2019roberta} for different programming languages. It is initially trained with general-purpose datasets and then is fine-tuned with downstream tasks, including code comment generation.

Some studies integrate structural information (AST) into tree-based encoders such as Graph neural network \cite{leclair2020improved}, Tree-LSTM \cite{shido2019automatic} and Tree-transformer \cite{harer2019tree}. Allamanis et al. \cite{allamanis2016convolutional} propose a convolutional  network to capture long-range attention knowledge and local time-invariant. 
In our work, we used flattened AST as the input to one of the encoders. However, our results show that AST downgrades the performance of Transformer slightly. 
The recent studies incorporate external knowledge to the neural model. 
Wei \cite{wei2019retrieve} and Zhang et al. \cite{zhang2020retrieval} incorporate information retrieval comment generation methods into neural machine translation models to improve the quality of the output comments. 
Given a source code snippet, they retrieve the most similar method where the corresponding comment is already available. Then, they exploit this comment as an additional input to the model to enhance the comment generation. 
Wang et al. \cite{wang2020cocogum} incorporate the external information to the model including enclosing class name and Unified Modeling Language. 
Haque et al. \cite{haque2020improved} integrate enclosing file context as the external knowledge into an attention mechanism to leverage words and tokens within the file to generate comments. They take the signature of other subroutines within the file into account and attempt to capture the relation between these subroutines and target subroutine summaries. 
The most similar work to our work is the approach of Hu et al. \cite{hu2018summarizing}. The authors use the name of the APIs to generate comments. Our work is different with existing works as it uses the API Docs for comment generation, which have not been explored previously. Also, our work is one of the few studies that leverages the Transformer model for comment generation, adding an external knowledge source to this model. 

There are many studies on comment generation and neural code representation, and in this section we provided a brief overview of the most relevant ones. For a survey on related work, please refer to the living literature review by Allamanis et al. \cite{allamanis2018survey}.

\section{Conclusion and Future Works} \label{sec:conclusion}

We presented the results of our model for code comment generation using API documentation as the external knowledge resource. The API Docs were extracted and concatenated to form as another input for each input code snippet. We performed experiments using both Transformer and RNN-based architectures. Interestingly, the results are not improved significantly. 
The results of more analysis shows that when number of API Docs increases, its benefit for the comment generation task decreases. We relate this to adding a long text (and therefore noise) to the input comment, which counteracts its benefits. 
Although the results are not promising in general, API Docs can be useful when the number of APIs are less than 3 in a method. These results can open avenues of research for new techniques to incorporate the API Docs for comment generation, such as adding weights to the API categories or identifying the informative APIs in a method.

\bibliographystyle{IEEEtran}
{

\footnotesize
\bibliography{    API2Com-RENE}

\begin{thebibliography}{10}
\providecommand{\url}[1]{#1}
\csname url@samestyle\endcsname
\providecommand{\newblock}{\relax}
\providecommand{\bibinfo}[2]{#2}
\providecommand{\BIBentrySTDinterwordspacing}{\spaceskip=0pt\relax}
\providecommand{\BIBentryALTinterwordstretchfactor}{4}
\providecommand{\BIBentryALTinterwordspacing}{\spaceskip=\fontdimen2\font plus
\BIBentryALTinterwordstretchfactor\fontdimen3\font minus
  \fontdimen4\font\relax}
\providecommand{\BIBforeignlanguage}[2]{{%
\expandafter\ifx\csname l@#1\endcsname\relax
\typeout{** WARNING: IEEEtran.bst: No hyphenation pattern has been}%
\typeout{** loaded for the language `#1'. Using the pattern for}%
\typeout{** the default language instead.}%
\else
\language=\csname l@#1\endcsname
\fi
#2}}
\providecommand{\BIBdecl}{\relax}
\BIBdecl

\bibitem{ko2006exploratory}
A.~J. Ko, B.~A. Myers, M.~J. Coblenz, and H.~H. Aung, ``An exploratory study of
  how developers seek, relate, and collect relevant information during software
  maintenance tasks,'' \emph{IEEE Transactions on software engineering},
  vol.~32, no.~12, pp. 971--987, 2006.

\bibitem{latoza2006maintaining}
T.~D. LaToza, G.~Venolia, and R.~DeLine, ``Maintaining mental models: a study
  of developer work habits,'' in \emph{Proceedings of the 28th international
  conference on Software engineering}, 2006, pp. 492--501.

\bibitem{sridhara2010towards}
G.~Sridhara, E.~Hill, D.~Muppaneni, L.~Pollock, and K.~Vijay-Shanker, ``Towards
  automatically generating summary comments for java methods,'' in
  \emph{Proceedings of the IEEE/ACM international conference on Automated
  software engineering}, 2010, pp. 43--52.

\bibitem{xia2017measuring}
X.~Xia, L.~Bao, D.~Lo, Z.~Xing, A.~E. Hassan, and S.~Li, ``Measuring program
  comprehension: A large-scale field study with professionals,'' \emph{IEEE
  Tran. on Software Eng.}, vol.~44, no.~10, pp. 951--976, 2017.

\bibitem{moreno2013automatic}
L.~Moreno, J.~Aponte, G.~Sridhara, A.~Marcus, L.~Pollock, and K.~Vijay-Shanker,
  ``Automatic generation of natural language summaries for java classes,'' in
  \emph{2013 21st International Conference on Program Comprehension
  (ICPC)}.\hskip 1em plus 0.5em minus 0.4em\relax IEEE, 2013, pp. 23--32.

\bibitem{singer2010examination}
J.~Singer, T.~Lethbridge, N.~Vinson, and N.~Anquetil, ``An examination of
  software engineering work practices,'' in \emph{CASCON First Decade High
  Impact Papers}, 2010, pp. 174--188.

\bibitem{eddy2013evaluating}
B.~P. Eddy, J.~A. Robinson, N.~A. Kraft, and J.~C. Carver, ``Evaluating source
  code summarization techniques: Replication and expansion,'' in \emph{2013
  21st ICPC}.\hskip 1em plus 0.5em minus 0.4em\relax IEEE, 2013, pp. 13--22.

\bibitem{haiduc2010use}
S.~Haiduc, J.~Aponte, L.~Moreno, and A.~Marcus, ``On the use of automated text
  summarization techniques for summarizing source code,'' in \emph{2010 17th
  Working Conf. on Reverse Eng.}\hskip 1em plus 0.5em minus 0.4em\relax IEEE,
  2010, pp. 35--44.

\bibitem{zhu2019automatic}
Y.~Zhu and M.~Pan, ``Automatic code summarization: A systematic literature
  review,'' \emph{arXiv preprint arXiv:1909.04352}, 2019.

\bibitem{iyer2016summarizing}
S.~Iyer, I.~Konstas, A.~Cheung, and L.~Zettlemoyer, ``Summarizing source code
  using a neural attention model,'' in \emph{Proceedings of the 54th Annual
  Meeting of the Association for Computational Linguistics (Volume 1: Long
  Papers)}, 2016, pp. 2073--2083.

\bibitem{hu2018deep}
X.~Hu, G.~Li, X.~Xia, D.~Lo, and Z.~Jin, ``Deep code comment generation,'' in
  \emph{2018 IEEE/ACM 26th International Conference on Program Comprehension
  (ICPC)}.\hskip 1em plus 0.5em minus 0.4em\relax IEEE, 2018, pp. 200--20\,010.

\bibitem{wan2018improving}
Y.~Wan, Z.~Zhao, M.~Yang, G.~Xu, H.~Ying, J.~Wu, and P.~S. Yu, ``Improving
  automatic source code summarization via deep reinforcement learning,'' in
  \emph{Proceedings of the 33rd ACM/IEEE International Conference on Automated
  Software Engineering}, 2018, pp. 397--407.

\bibitem{movshovitz2013natural}
D.~Movshovitz-Attias and W.~Cohen, ``Natural language models for predicting
  programming comments,'' in \emph{Proceedings of the 51st Annual Meeting of
  the Association for Computational Linguistics (Volume 2: Short Papers)},
  2013, pp. 35--40.

\bibitem{bahdanau2014neural}
D.~Bahdanau, K.~Cho, and Y.~Bengio, ``Neural machine translation by jointly
  learning to align and translate,'' \emph{arXiv preprint:1409.0473}, 2014.

\bibitem{haije2016automatic}
T.~Haije, B.~O.~K. Intelligentie, E.~Gavves, and H.~Heuer, ``Automatic comment
  generation using a neural translation model,'' \emph{Inf. Softw. Technol},
  vol.~55, no.~3, pp. 258--268, 2016.

\bibitem{leclair2019neural}
A.~LeClair, S.~Jiang, and C.~McMillan, ``A neural model for generating natural
  language summaries of program subroutines,'' in \emph{2019 IEEE/ACM 41st
  International Conference on Software Engineering (ICSE)}.\hskip 1em plus
  0.5em minus 0.4em\relax IEEE, 2019, pp. 795--806.

\bibitem{wei2019retrieve}
B.~Wei, ``Retrieve and refine: exemplar-based neural comment generation,'' in
  \emph{2019 34th IEEE/ACM International Conference on Automated Software
  Engineering (ASE)}.\hskip 1em plus 0.5em minus 0.4em\relax IEEE, 2019, pp.
  1250--1252.

\bibitem{zhang2020retrieval}
J.~Zhang, X.~Wang, H.~Zhang, H.~Sun, and X.~Liu, ``Retrieval-based neural
  source code summarization,'' in \emph{Proceedings of the 42nd International
  Conference on Software Engineering. IEEE}, 2020.

\bibitem{wang2020cocogum}
\BIBentryALTinterwordspacing
Y.~Wang, L.~Du, E.~Shi, Y.~Hu, S.~Han, and D.~Zhang, ``Cocogum: Contextual code
  summarization with multi-relational gnn on umls,'' Microsoft, Tech. Rep.
  MSR-TR-2020-16, May 2020. [Online]. Available:
  \url{https://www.microsoft.com/en-us/research/publication/cocogum-contextual-code-summarization-with-multi-relational-gnn-on-umls/}
\BIBentrySTDinterwordspacing

\bibitem{haque2020improved}
S.~Haque, A.~LeClair, L.~Wu, and C.~McMillan, ``Improved automatic
  summarization of subroutines via attention to file context,'' in
  \emph{Proceedings of the 17th International Conference on Mining Software
  Repositories}, 2020, pp. 300--310.

\bibitem{hu2018summarizing}
X.~Hu, G.~Li, X.~Xia, D.~Lo, S.~Lu, and Z.~Jin, ``Summarizing source code with
  transferred api knowledge,'' 2018.

\bibitem{alon2018code2seq}
U.~Alon, S.~Brody, O.~Levy, and E.~Yahav, ``code2seq: Generating sequences from
  structured representations of code,'' \emph{arXiv preprint arXiv:1808.01400},
  2018.

\bibitem{vaswani2017attention}
A.~Vaswani, N.~Shazeer, N.~Parmar, J.~Uszkoreit, L.~Jones, A.~N. Gomez,
  {\L}.~Kaiser, and I.~Polosukhin, ``Attention is all you need,'' in
  \emph{Advances in neural information processing systems}, 2017, pp.
  5998--6008.

\bibitem{husain2019codesearchnet}
H.~Husain, H.-H. Wu, T.~Gazit, M.~Allamanis, and M.~Brockschmidt,
  ``Codesearchnet challenge: Evaluating the state of semantic code search,''
  \emph{arXiv preprint arXiv:1909.09436}, 2019.

\bibitem{chung2014empirical}
J.~Chung, C.~Gulcehre, K.~Cho, and Y.~Bengio, ``Empirical evaluation of gated
  recurrent neural networks on sequence modeling,'' \emph{arXiv preprint
  arXiv:1412.3555}, 2014.

\bibitem{hochreiter1997long}
S.~Hochreiter and J.~Schmidhuber, ``Long short-term memory,'' \emph{Neural
  computation}, vol.~9, no.~8, pp. 1735--1780, 1997.

\bibitem{ahmad2020transformer}
W.~U. Ahmad, S.~Chakraborty, B.~Ray, and K.-W. Chang, ``A transformer-based
  approach for source code summarization,'' \emph{arXiv preprint
  arXiv:2005.00653}, 2020.

\bibitem{wang2019learning}
Q.~Wang, B.~Li, T.~Xiao, J.~Zhu, C.~Li, D.~F. Wong, and L.~S. Chao, ``Learning
  deep transformer models for machine translation,'' \emph{arXiv preprint
  arXiv:1906.01787}, 2019.

\bibitem{you2019improving}
Y.~You, W.~Jia, T.~Liu, and W.~Yang, ``Improving abstractive document
  summarization with salient information modeling,'' in \emph{Proceedings of
  the 57th Annual Meeting of the Association for Computational Linguistics},
  2019, pp. 2132--2141.

\bibitem{fan2018hierarchical}
A.~Fan, M.~Lewis, and Y.~Dauphin, ``Hierarchical neural story generation,''
  \emph{arXiv preprint arXiv:1805.04833}, 2018.

\bibitem{wang2020trans}
W.~Wang, Y.~Zhang, Z.~Zeng, and G.~Xu, ``Trans\^{} 3: A transformer-based
  framework for unifying code summarization and code search,'' \emph{arXiv
  preprint arXiv:2003.03238}, 2020.

\bibitem{shaw-etal-2018-self}
\BIBentryALTinterwordspacing
P.~Shaw, J.~Uszkoreit, and A.~Vaswani, ``Self-attention with relative position
  representations,'' in \emph{Proceedings of the 2018 Conference of the North
  {A}merican Chapter of the Association for Computational Linguistics: Human
  Language Technologies, Volume 2 (Short Papers)}.\hskip 1em plus 0.5em minus
  0.4em\relax New Orleans, Louisiana: Association for Computational
  Linguistics, Jun. 2018, pp. 464--468. [Online]. Available:
  \url{https://www.aclweb.org/anthology/N18-2074}
\BIBentrySTDinterwordspacing

\bibitem{feng2020codebert}
Z.~Feng, D.~Guo, D.~Tang, N.~Duan, X.~Feng, M.~Gong, L.~Shou, B.~Qin, T.~Liu,
  D.~Jiang \emph{et~al.}, ``Codebert: A pre-trained model for programming and
  natural languages,'' \emph{arXiv preprint arXiv:2002.08155}, 2020.

\bibitem{papineni2002bleu}
K.~Papineni, S.~Roukos, T.~Ward, and W.-J. Zhu, ``Bleu: a method for automatic
  evaluation of machine translation,'' in \emph{Proc. of the 40th annual
  meeting of the Assoc. for Computational Linguistics}, 2002, pp. 311--318.

\bibitem{lin2004rouge}
C.-Y. Lin, ``Rouge: A package for automatic evaluation of summaries,'' in
  \emph{Text summarization branches out}, 2004, pp. 74--81.

\bibitem{banerjee2005meteor}
S.~Banerjee and A.~Lavie, ``Meteor: An automatic metric for mt evaluation with
  improved correlation with human judgments,'' in \emph{Proceedings of the acl
  workshop on intrinsic and extrinsic evaluation measures for machine
  translation and/or summarization}, 2005, pp. 65--72.

\bibitem{hu2020deep}
X.~Hu, G.~Li, X.~Xia, D.~Lo, and Z.~Jin, ``Deep code comment generation with
  hybrid lexical and syntactical information,'' \emph{Empirical Software
  Engineering}, vol.~25, no.~3, pp. 2179--2217, 2020.

\bibitem{stapleton2020human}
S.~Stapleton, Y.~Gambhir, A.~LeClair, Z.~Eberhart, W.~Weimer, K.~Leach, and
  Y.~Huang, ``A human study of comprehension and code summarization,'' in
  \emph{Proc. of the 28th ICPC}, 2020, pp. 2--13.

\bibitem{mcburney2015automatic}
P.~W. McBurney and C.~McMillan, ``Automatic source code summarization of
  context for java methods,'' \emph{IEEE Transactions on Software Engineering},
  vol.~42, no.~2, pp. 103--119, 2015.

\bibitem{wong2013autocomment}
E.~Wong, J.~Yang, and L.~Tan, ``Autocomment: Mining question and answer sites
  for automatic comment generation,'' in \emph{2013 28th IEEE/ACM International
  Conference on Automated Software Engineering (ASE)}.\hskip 1em plus 0.5em
  minus 0.4em\relax IEEE, 2013, pp. 562--567.

\bibitem{wei2019code}
B.~Wei, G.~Li, X.~Xia, Z.~Fu, and Z.~Jin, ``Code generation as a dual task of
  code summarization,'' in \emph{Advances in Neural Information Processing
  Systems}, 2019, pp. 6563--6573.

\bibitem{devlin2018bert}
J.~Devlin, M.-W. Chang, K.~Lee, and K.~Toutanova, ``Bert: Pre-training of deep
  bidirectional transformers for language understanding,'' \emph{arXiv preprint
  arXiv:1810.04805}, 2018.

\bibitem{liu2019roberta}
Y.~Liu, M.~Ott, N.~Goyal, J.~Du, M.~Joshi, D.~Chen, O.~Levy, M.~Lewis,
  L.~Zettlemoyer, and V.~Stoyanov, ``Roberta: A robustly optimized bert
  pretraining approach,'' \emph{arXiv preprint arXiv:1907.11692}, 2019.

\bibitem{leclair2020improved}
A.~LeClair, S.~Haque, L.~Wu, and C.~McMillan, ``Improved code summarization via
  a graph neural network,'' in \emph{Proc. of the 28th Int. Conference on
  Program Comprehension}, 2020, pp. 184--195.

\bibitem{shido2019automatic}
Y.~Shido, Y.~Kobayashi, A.~Yamamoto, A.~Miyamoto, and T.~Matsumura, ``Automatic
  source code summarization with extended tree-lstm,'' in \emph{2019 Int. Joint
  Conf. on Neural Networks (IJCNN)}.\hskip 1em plus 0.5em minus 0.4em\relax
  IEEE, 2019, pp. 1--8.

\bibitem{harer2019tree}
J.~Harer, C.~Reale, and P.~Chin, ``Tree-transformer: A transformer-based method
  for correction of tree-structured data,'' \emph{arXiv preprint
  arXiv:1908.00449}, 2019.

\bibitem{allamanis2016convolutional}
M.~Allamanis, H.~Peng, and C.~Sutton, ``A convolutional attention network for
  extreme summarization of source code,'' in \emph{International conference on
  machine learning}.\hskip 1em plus 0.5em minus 0.4em\relax PMLR, 2016, pp.
  2091--2100.

\bibitem{allamanis2018survey}
M.~Allamanis, E.~T. Barr, P.~Devanbu, and C.~Sutton, ``A survey of machine
  learning for big code and naturalness,'' \emph{ACM Computing Surveys (CSUR)},
  vol.~51, no.~4, p.~81, 2018.

\end{thebibliography}

}
\end{document}